\documentclass[manuscipt]{aastex631}
\usepackage{placeins}
\usepackage{amsmath}
\usepackage{graphicx}

\received{December 9, 2023}
\accepted{December  22, 2023}

\submitjournal{\apjl}

\shorttitle{New Horizons Venetia Burney Student Dust Counter (VBSDC)}
\shortauthors{Doner et al.}

\begin{document}

\title{New Horizons Venetia Burney Student Dust Counter Observes Higher than Expected Fluxes Approaching 60 AU}
\correspondingauthor{Alex Doner}
\email{alex.doner@lasp.colorado.edu}

\author[0000-0001-7065-3224]{Alex Doner}
\affiliation{Laboratory for Atmospheric and Space Physics,  University of Colorado, Boulder, CO 80303, USA}
\affiliation{Department of Physics,  University of Colorado, Boulder, CO, USA}

\author[0000-0002-5920-9226]{Mih\'aly Hor\'anyi}
\affiliation{Laboratory for Atmospheric and Space Physics,  University of Colorado, Boulder, CO 80303, USA}
\affiliation{Department of Physics,  University of Colorado, Boulder, CO, USA}

\author[0000-0002-3963-1614]{Fran Bagenal}
\affiliation{Laboratory for Atmospheric and Space Physics,  University of Colorado, Boulder, CO 80303, USA}

\author[0000-0002-4644-0306]{Pontus Brandt}
\affiliation{The Johns Hopkins University Applied Physics Laboratory, Laurel, MD, USA}

\author[0000-0002-8296-6540]{Will Grundy}
\affiliation{Lowell Observatory, Flagstaff, AZ, 86001, USA}

\author[0000-0002-9548-1526]{Carey Lisse}
\affiliation{The Johns Hopkins University Applied Physics Laboratory, Laurel, MD, USA}

\author[0000-0002-3672-0603]{Joel Parker}
\affiliation{Southwest Research Institute, Boulder, CO, USA}

\author[0000-0001-8137-8176]{Andrew R. Poppe}
\affiliation{Space Sciences Laboratory, University of California, Berkeley, CA, USA}

\author[0000-0003-3045-8445]{Kelsi N. Singer}
\affiliation{Southwest Research Institute, Boulder, CO, USA}

\author[0000-0001-5018-7537]{S. Alan Stern}
\affiliation{Southwest Research Institute, Boulder, CO, USA}

\author[0000-0002-3323-9304]{Anne Verbiscer}
\affiliation{Department of Astronomy, University of Virginia, Charlottesville, VA, 22904, USA}

\begin{abstract}

The NASA New Horizons Venetia Burney Student Dust Counter (SDC) measures dust particle impacts along the spacecraft's flight path for grains with mass $\ge$ $10^{-12}$ g, mapping out their spatial density distribution. We present the latest SDC dust density, size distribution, and flux measurements through 55 au and compare them to numerical model predictions. Kuiper Belt Objects (KBOs) are thought to be the dominant source of interplanetary dust particles (IDP) in the outer solar system due to both collisions between KBOs, and their continual bombardment by interstellar dust particles (ISD).
 Continued measurements through 55 au show higher than model-predicted dust fluxes as New Horizons approaches the putative outer edge of the Kuiper Belt (KB). We discuss potential explanations for the growing deviation: radiation pressure stretches the dust distribution to further heliocentric distances than its parent body distribution; icy dust grains undergo photo-sputtering that rapidly increases their response to radiation pressure forces and pushes them further away from the sun; and the distribution of KBOs may extend much further than existing observations suggest. Ongoing SDC measurements at even larger heliocentric distances will continue to constrain the contributions of dust production in the KB. Continued SDC measurements remain crucial for understanding the Kuiper Belt and the interpretation of observations of dust disks around other stars.

\end{abstract}


%%%%%%%%%%%%%%%%%%%%
\section{Introduction} 
\label{sec:intro}

\indent Interplanetary dust particles (IDPs) carry information about the birth and evolution of planetary bodies within our solar system. The orbital distribution of IDPs is driven by the forces of gravity from the Sun and planets, radiation pressure, Poynting-Robertson (PR) drag, and electromagnetic forces \citep{liou:96a, malhotra:02a, malhotra:03a, poppe:16c}. Additionally, the size and mass of individual IDPs are constantly evolving due to sputtering and mutual collisions. While  Jupiter Family Comets (JFCs) and asteroids dominate IDP production in the inner solar system \citep{spiesman:95, kelsall:98a, nesvorny:10a}, the Kuiper Belt (KB) is the dominant source of IDPs in the outer solar system \citep{stern:96a,yamamoto:98a} from either mutual Kuiper Belt Object (KBO) collisions or interstellar dust bombardment. While IDPs created outside 30 au generally flow towards the Sun, Neptune typically prolongs their lifetime due to resonances \citep{liou:96a, malhotra:02a, kuchner:10a}.

The Venetia Burney Student Dust Counter (SDC) \citep{horanyi:08a} is the first dedicated instrument to make in situ dust measurements beyond 17 au from the Sun. This paper reports the most recent SDC observations, similar to our earlier reports summarizing the measurements as the mission has traversed across our solar system \citep{poppe:10a, han:11a, szalay:13a, piquette:19a, bernardoni:22a}. We also offer an as assessment of the processes that might be responsible for the higher than expected dust fluxes reported in this paper.

%%%%%%%%%%%%%%%%%%%%
\section{Instrument Description} \label{sec:instrument}

\indent SDC is composed of 14 permanently electrically polarized polyvinylidene fluoride (PVDF) plastic film detectors \citep{horanyi:08a}. Each film is 28 $\mu$m thick and has dimensions of 14.2 x 6.5 cm. The SDC detector panel is mounted on the ram side of the spacecraft such that the front  12 detectors are exposed to dust impacts and 2 reference detectors remain shielded from dust and are used as control detectors to identify spurious noise events. PVDF impact detectors operate by measuring a change in the surface charge density on the conducting surface coatings of the films from craters formed by dust impacts. For each impact, the peak measured charge is a function of both the impactor's mass, $m$, and velocity, $v$ \citep{piquette:20a}.

PVDF detectors have piezoelectric and pyroelectric properties that make them responsive to thermal fluctuations and vibrations on the spacecraft. The 2 shielded detectors provide a baseline for non-dust noise events and serve as references for the 12 science detectors. The detectors are further grouped into two channels of 6 science detectors and 1 reference detector, labeled Channels A and B, where each channel has its own electronics chain and analog-to-digital converter (ADC). \\
\indent The instrument undergoes periodic noise and stimulus tests to monitor any possible sensitivity degradation. For each detector, noise tests measure the rate of impacts that are above successive threshold values. The thresholds initially start below typical operational values and are gradually stepped up throughout the test. This produces an array of impact rates per threshold level per detector. The results are used to update the optimal threshold levels, especially in preparation for periods of high spacecraft activity with intense periods of spacecraft generated noise that can mimic dust impacts, such as the encounter with Pluto \citep{stern:15a,bagenal:16a}. In addition, 
charge stimulus tests are also used to observe any changes in the calibration of the electronics by injecting a series of known charges from capacitors into each detector's electronics chain. These tests show that the instrument has remained stable since its launch in 2006, and there are no aging-related effects that would require changing the analysis and interpretation of its data \citep{fountain:23a}.

%%%%%%%%%%%%%%%%%%%%
\begin{figure}[t]
    \begin{center}
	\includegraphics[width=0.9\textwidth]{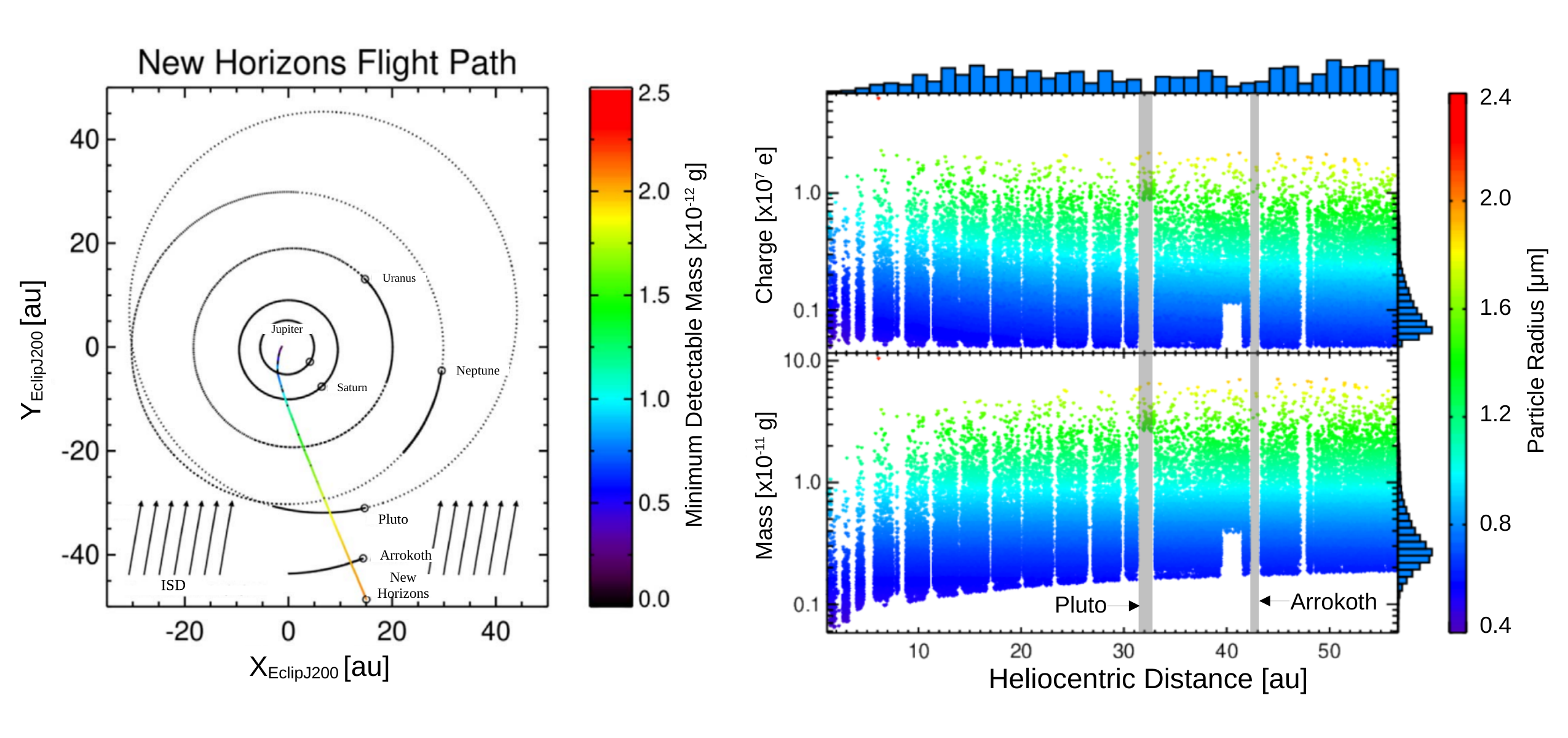}
   \end{center}
    \caption{
\textbf{Left:} The flight path of the New Horizons spacecraft up through 55 au with the minimum detectable mass represented in the color scale. The decrease in spacecraft speed with increasing distance results in a higher minimum detectable mass. New Horizons is heading along the ecliptic longitude $\lambda_{\rm NH}  \simeq$ 293° compared to the ISD inflow of $\lambda_{\rm ISD}$ = 259°, indicated by the parallel upward-pointing arrows at the bottom. Therefore, ISDs impact SDC at an angle of $\alpha\simeq$ 23° off of normal \citep{bernardoni:22a}. ISD's currently impact SDC at 39 km/s while IDP's only impact at 14 km/s. Therefore, despite their size difference, ISD impacts can be measured and interpreted as small IDPs \citep{bernardoni:22a}. Based on the current model, ISD assumed IDP size should not exceed 0.68 $\mu$m.
\textbf{Right:} All non-coincident data up to July 25th, 2023. Each detector has an independent mass threshold. For flux and density computation a common threshold was set at $r_g \ge 0.63 ~\mu$m ($m \ge 2.62 \times 10^{-12}$ g) such that all of the detectors provide an equal statistical contribution to the calculated flux and density \citep{bernardoni:22a}.
The grey blocks represent regions of exceptionally high thresholds due to high spacecraft activity during close encounters with Pluto and Arrokoth. Other periods of missing data are due to SDC being powered off for spacecraft operational reasons.
}
\label{flight_path_charge}
\end{figure}

%%%%%%%%%%%%%%%%%%%%
\section{Flux and density measurements}

\indent SDC has been monitoring dust impacts nearly continuously since the New Horizons launch on January 19th, 2006, and it is now measuring dust beyond 55 au. Figure \ref{flight_path_charge}(Left), shows the up-to-date trajectory of the New Horizons spacecraft through the solar system, as well as the corresponding minimum detectable mass given the assumption that SDC is measuring predominantly silicate dust grains with a density of 2.5 g/cm$^3$. The SDC impact velocity is derived from the velocity difference between the spacecraft and IDP grains that are assumed to be on circular Kepler orbits that are modified by radiation pressure. As the spacecraft gets further away from the Sun and slows down, the particle mass required to generate the same charge magnitude increases thus increasing the minimum detection mass threshold. The circular dust orbit assumption is a source of uncertainty as the dust may be on eccentric and/or inclined orbits which alter the true impact speed. Thus, there is also an introduced uncertainty in the detected mass and minimum detectable mass.

\begin{figure}[h]
\hskip 1.5cm
\vbox{
  \hbox{
  \includegraphics[width=0.4\linewidth]{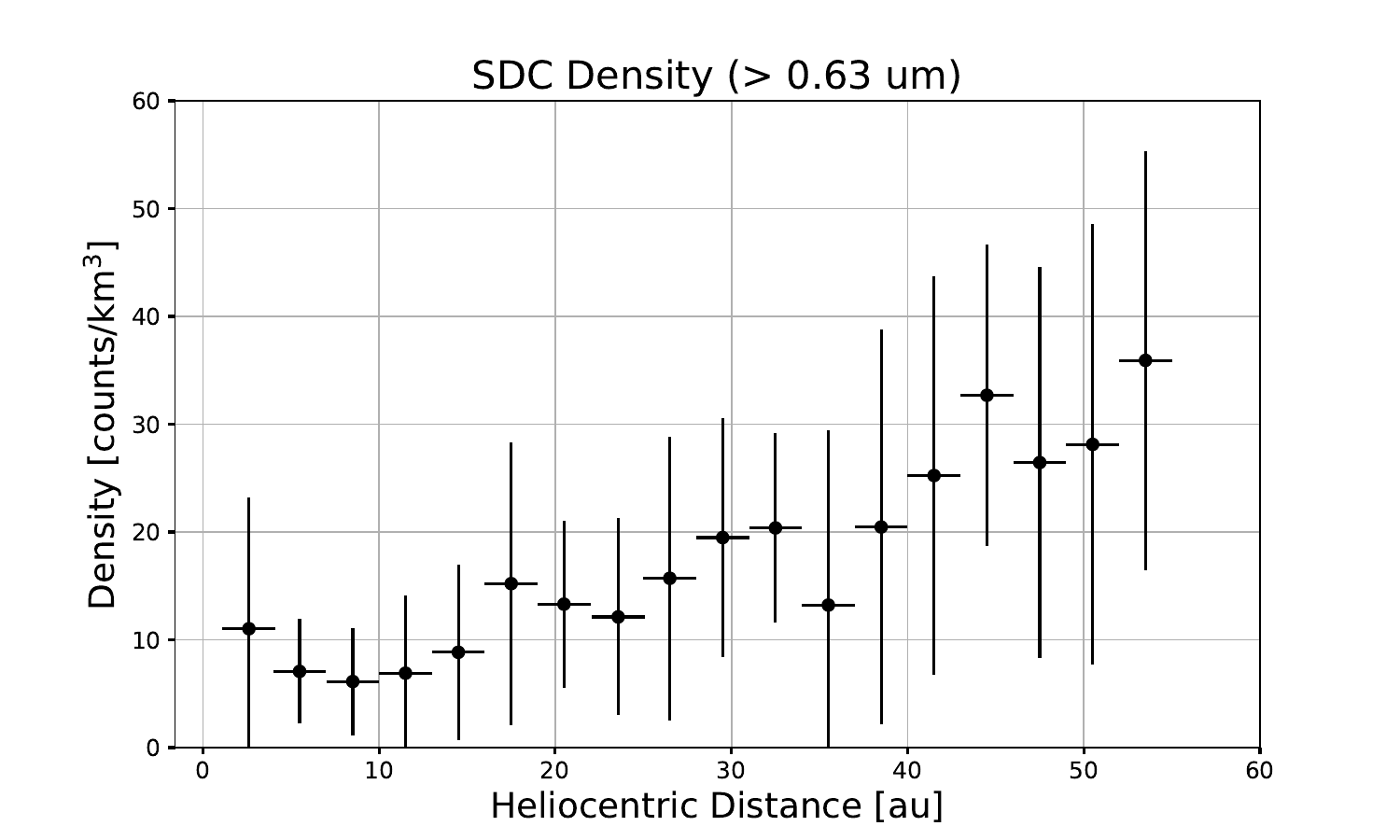}
  \includegraphics[width=0.4\linewidth]{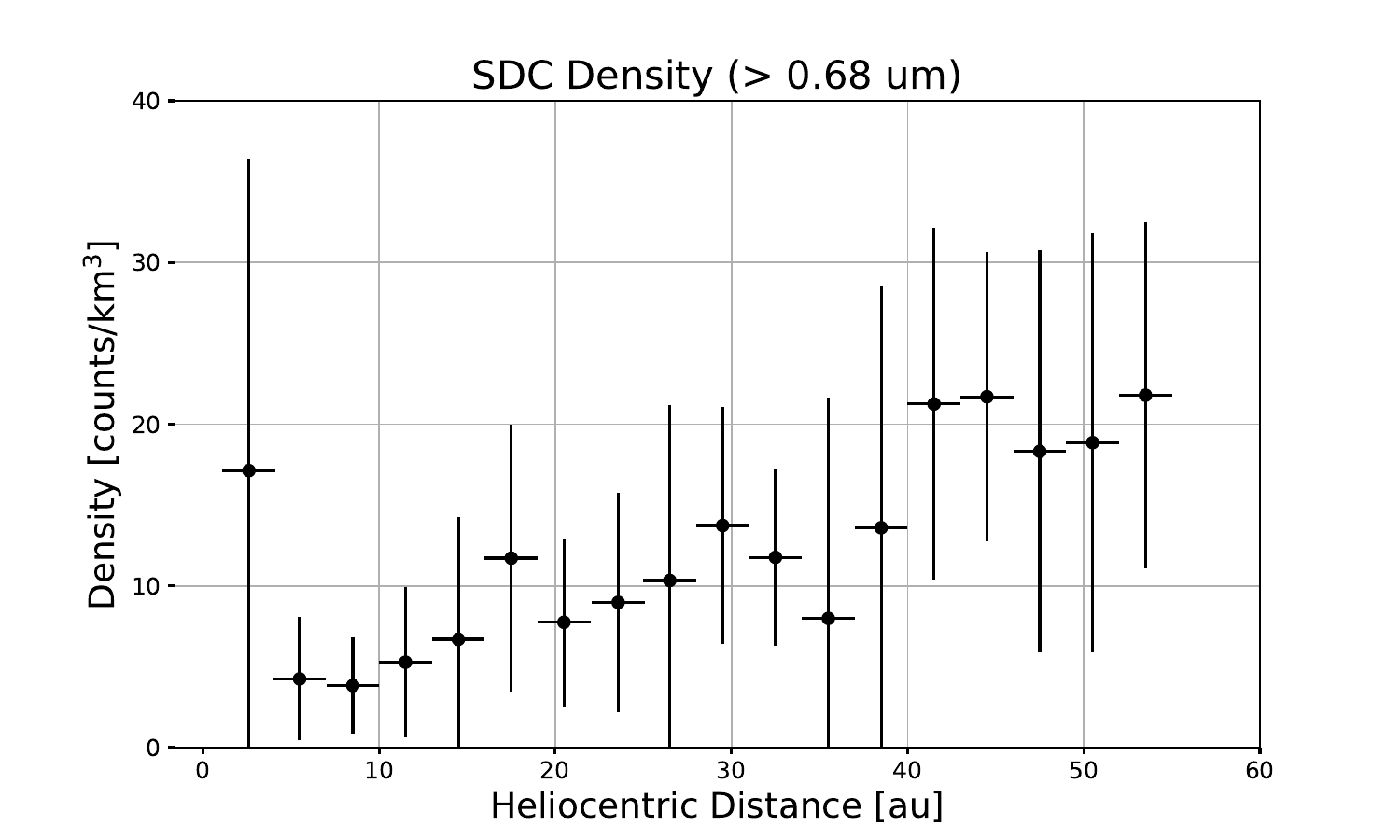}}
     \hbox{
     \includegraphics[width=0.4\linewidth]
      {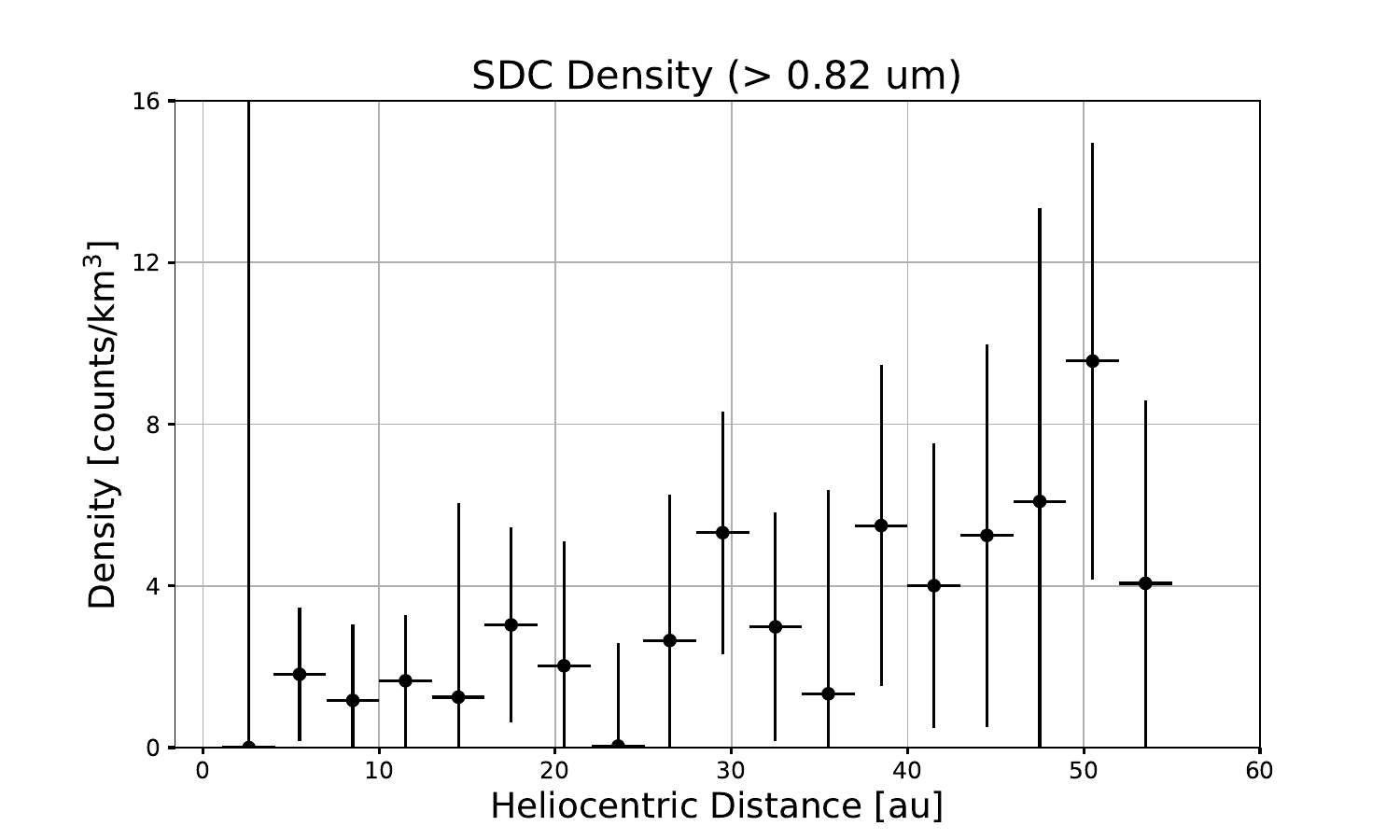}
     \includegraphics[width=0.4\linewidth]{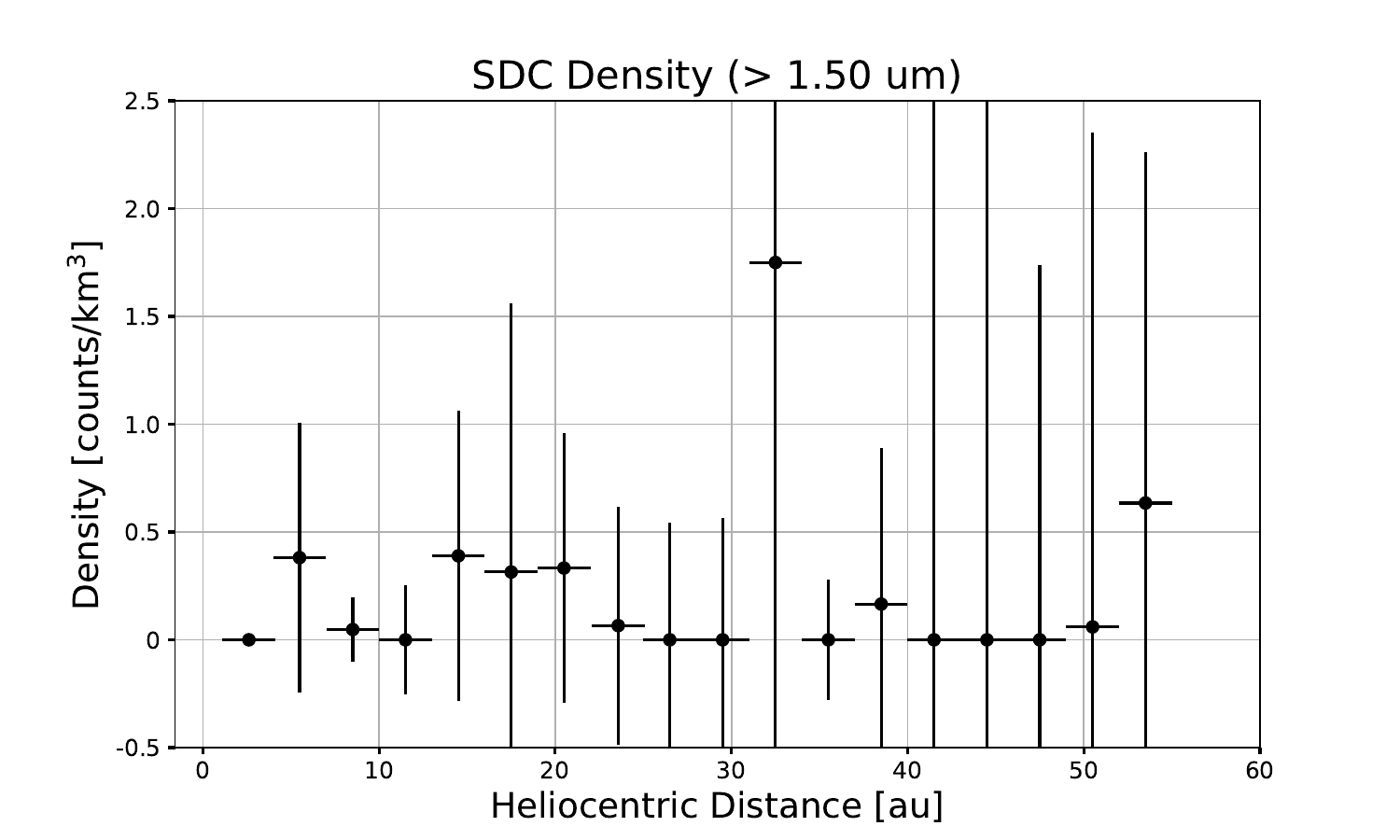}
}}
  \caption{SDC density measurements above 4 cutoff sizes: 0.63 $\mu$m, 0.68 $\mu$m, 0.82 $\mu$m, and 1.50 $\mu$m. The overall shape and upward trend, within error bars, is consistent between 0.63, 0.68, and 0.82 $\mu$m. Above 1.50 $\mu$m, there are insufficient detections to make any confident claims about the density of the grains throughout the New Horizons flight path. The 0.63, 0.82 and 1.50 $\mu$m cutoffs are chosen based on the on-board threshold table for different flight conditions, while the 0.68 $\mu$m cutoff is selected to observe the flux without any potential effects from interstellar grains.} 
  \label{density}
\end{figure}

\indent Due to possible noise events, measuring only the number and magnitude of potential impacts, is insufficient to correctly calculate the flux and density of interplanetary dust along the spacecraft's flight path. All events that are recorded within 1-10 seconds of another detector's event, due to thruster firings or mechanical vibrations, are flagged as \textit{coincident} and removed from the science event list. After this data filtering is applied, the final data set can be seen in Figure \ref{flight_path_charge} showing the direct charge measurements as a function of the spacecraft's heliocentric distance, and the corresponding mass for IDP's within our assumptions \citep{piquette:19a,bernardoni:22a}. Before computing density or flux, due to the independent thresholds of each detector throughout the operation of the instrument, a common minimum size cutoff is set at $ r_g \ge 0.63$ ~$\mu$m such that each detector's measurements may contribute equally. The average spatial dust density is then estimated for each detector's signals with $r_g \ge 0.63$ $\mu$m in 3 au traversed bins. This is computed by dividing the number of counts by the swept volume for the time that the detector is turned on with thresholds below the $ r_g = 0.63 ~\mu$m cutoff. The swept volume of each detector is defined by \citep{piquette:19a}:
\begin{equation}
    V = A_{det} \int_{T1}^{T2} \mathbf{\hat{n}}_{SDC} \cdot (\mathbf{v}_{sc} - \mathbf{v}_{dust}) dt,
\end{equation}
where $A_{det}$ is the detector area, $\mathbf{\hat{n}}_{SDC}$ is the pointing direction of SDC, $\mathbf{v}_{sc}$ is the spacecraft velocity, and $\mathbf{v}_{dust}$ is the dust velocity, assuming the dust is on circular Kepler orbits modified by radiation pressure.
The events recorded on the reference detectors are identified as noise and subtracted from the density measurement of each science detector that shares its corresponding electronics chain. The average spatial dust density across the science detectors is calculated in 3 au bins. The 3 au bin size presented the best balance between minimizing uncertainty and preserving spatial features. Figure \ref{density} shows the average spatial density estimates in 3 au bins for four different size cutoffs: $r_g \ge$  0.63;\ 0.68;\ 0.82; and\ 1.5\ $\mu m$.

\begin{figure}[h]
    \begin{center}
	\includegraphics[width=0.7\textwidth]{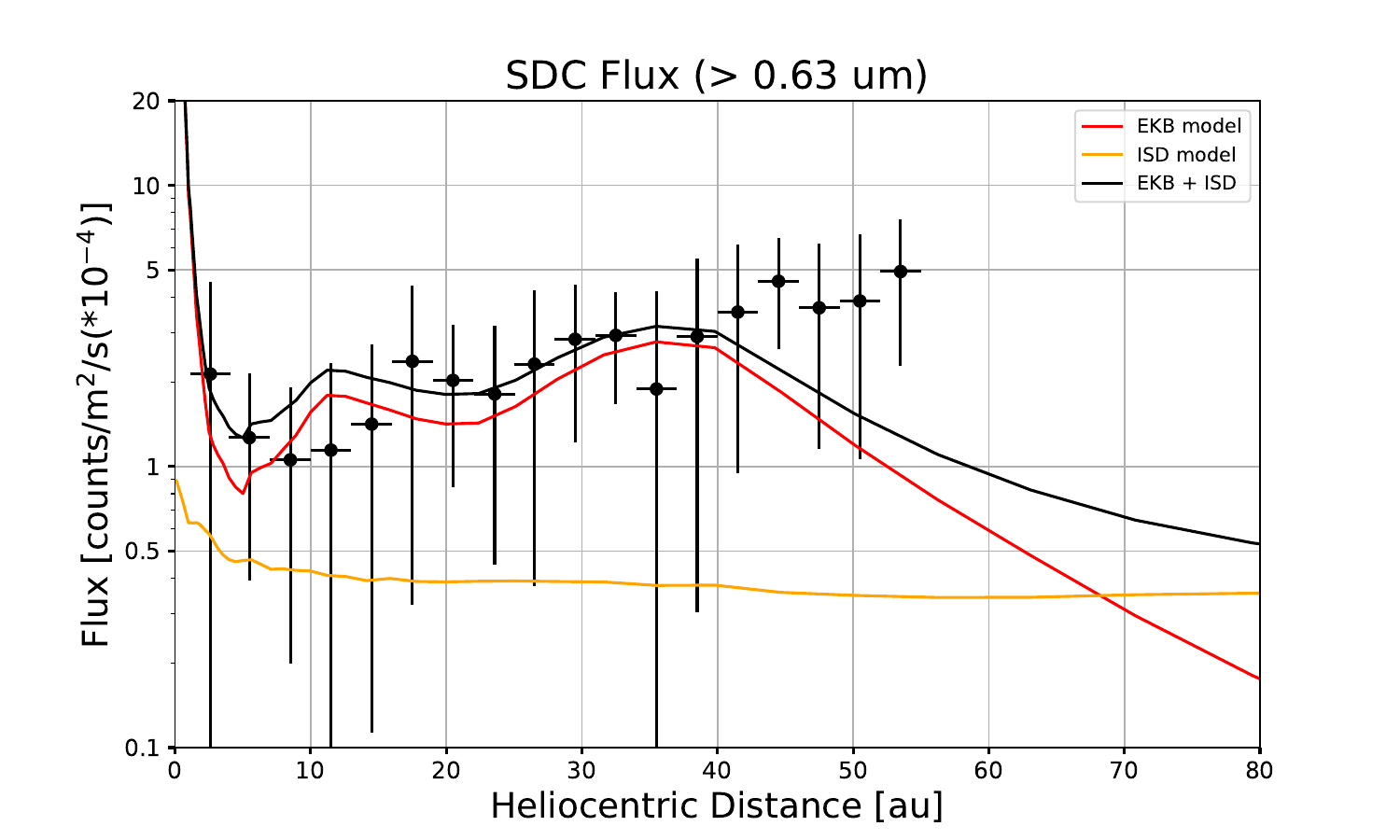}
    \end{center}
    \caption{
	SDC flux estimates for particles with a radius greater than 0.63 $\mu$m from 1 to 55 heliocentric au. Each point, with 1-$\sigma$ error bars, is an average of the flux measured by each detector across each 3 au traversed by the New Horizons spacecraft. The orange curve represents the ISD flux values calibrated from Ulysses measurements \citep{bernardoni:22a}. While ISD grains are much smaller ($\sim0.28 \mu$m), they are also traveling at higher relative velocities and, as a result, SDC interprets them as small IDPs \citep{bernardoni:22a}. The red line shows a model from \citep{poppe:16c} that is normalized to the SDC measurements and assumes SDC is impacted only by IDPs. The black line is a sum of the orange and red models to provide a more representative model of the total flux that SDC is measuring up to 42 au. 
 }
\label{flux}
\end{figure}

The flux for each detector is computed by dividing the swept volume by the spacecraft's traversed distance while the detector was on and dividing by the detector's total on time. As with the density, the apparent flux of each reference detector is subtracted from each of the science detectors on their respective electronics chains and the average flux across the science detectors for each 3 au bin is reported. Figure \ref{flux} compares the SDC flux estimate to models based on earlier SDC reports \citep{poppe:16c, bernardoni:22a}. Beyond 42 au, the data reveals a growing divergence from the models, potential explanations for which are discussed in section \ref{sec:dust}.

%%%%%%%%%%%%%%%%%%%%

\FloatBarrier
%%%%%%%%%%%%%%%%%%%%
\section{Dust production and dynamics in the Kuiper belt} \label{sec:dust}
To gain insight into the possible interpretations of the higher-than-expected dust fluxes reported by SDC beyond 42 au, we revisit the basic ideas of dust production in the KB \citep{moro-martin:08a, liou:08a, poppe:16c} and in dust disks around other stars \citep{krivov:07a, krivov:10a, grigorieva:07a}. Each KBO is a source of dust particles due to collisions with other KBOs \citep{stern:96a} and due to  their bombardment by interstellar dust particles \citep{yamamoto:98a}. In either case, the ejection velocity of the produced dust particles remains small compared to the orbital speeds of their parent bodies, so they are born with approximately the same state vectors as their parent bodies; however, these dust particles have a hugely different $\beta$'s -- where $\beta$ is the ratio of the repulsive radiation pressure to the attractive gravity on the particle \citep{kresak:76a, burns:79a}. The result is a near-instantaneous change in their specific (i.e., per unit mass) potential energy while keeping their parent body's specific kinetic energy and angular momentum. Due to the apparent reduction of solar gravity by a factor of $(1-\beta)$, their total specific energy and angular momentum after their release are
\begin{eqnarray}
    E^{*} & = & \frac{1}{2}({\dot r}^2 + r^2{\dot \theta}^2) - \frac{(1-\beta) \mu}{r}  =  E + \frac{\beta\mu}{r}, \rm{\ \ and} \label{eq:E} \\
    J^{*} & = & r^2{\dot \theta} =  J, \label{eq:J}    
\end{eqnarray}
where $r$ and $\theta$ are polar coordinates in the coordinate system centered on the Sun;  $\mu = GM_{\odot}$ is the product of the gravitational constant and the solar mass; $E$ and $J$ are the total specific energy and angular momentum before the release. From equation \ref{eq:E}, exploiting the relationship  between the semi-major axis,  $a$,  of an orbit and its total energy,  $ a = -\mu / (2 E)$,  the new semi-major axis of a dust particle, $a^*$, can be calculated \citep{kresak:76a, liou:08a}
\begin{equation}
    a^* = \frac{(1-\beta)  a r}{r - 2\beta a}.
    \label{eq:a}
\end{equation}

Similarly, using the relationship between the orbital angular momentum, semi-major axis, and eccentricity, $e$,  ($ J = \sqrt{ \mu a (1 - e^2)}$, the new orbit's eccentricity becomes

\begin{equation}
    e^* = \sqrt{
    1 - \frac{(1 - e^2)(r - 2 \beta a )}
{(1 - \beta)^2 r}}.
\label{eq:e}
\end{equation}

\begin{figure}
    \centering
    \includegraphics[width=\textwidth]{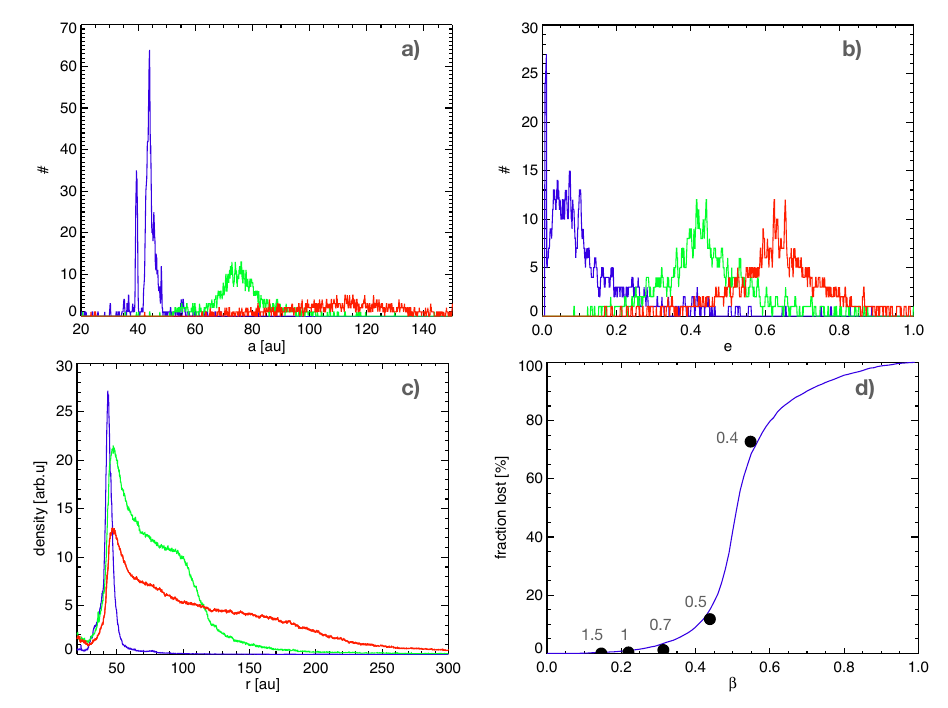}
    \caption{\textbf{a)} The semi-major axis, $a$, distribution of the low-inclination ($i \le 10^{\circ}$) KBOs (blue line), and of the newly released particles, $a^*$, for $\beta$ = 0.3 (green) and 0.4 (red). \textbf{b)} The eccentricity distributions of parent bodies, $e$, and the new dust particles, $e^*$. \textbf{c)} the radial density distributions, each curve normalized to the total number of KBOs (2320) or dust particles (2290 and 2200). \textbf{d)} The fraction of the ejected particles generated from all low-inclination KBOs as a function of $\beta$. The dots show the results of a full numerical solution of the equation of motion for various radii (measured in $\mu$m) olivine-type particles \citep{poppe:16c}.}
    \label{fig:Kepler}
\end{figure}

As a function of the parent body's orbital elements and position along its orbit when a dust particle is released, as well as  the composition and size of the released particle ($\beta$),  equations \ref{eq:E} and \ref{eq:J} can result in $a^*\le 0$  and $e^*\ge 1$, indicating that the particle will not follow a bound orbit and escapes from the solar system.  
The combination of higher $\beta$ values and a release point closer to the pericenter of the parent body results in more escaping particles. Particles that remain on bound orbits will form a dust disk that extends much beyond the spatial distribution of their parent bodies. These phenomena must hold for dust disks around other stars as well. 

There are currently $\sim$5200 KBOs with estimates of their orbital elements (Minor Planet Center, October, 2023). We downselected for those with inclinations $i \le 10^{\circ}$, leaving 2320 KBOs that are expected to dominate the dust production that contributes to SDC measurements near the ecliptic plane. The probability of finding one of the KBOs with orbital elements $a_i, e_i$  at a distance $r$ is 
\begin{equation}
    p_i(r) = \begin{cases} 
    \frac{C_i}{\sqrt{a_i^2 e_i^2 - (r -a_i)^2}} &  \text{if } \ \ a_i (1 -e_i) \le r \le a_i (1+e_i),  \\
                                       0  & \text{otherwise, }
    \end{cases}  
\end{equation}
where $C_i$ is a normalization constant to set $\int_{0}^{\infty} p_i(r) dr = 1$ \citep{horanyi:92a}. The semi-major axis and eccentricity distributions, as well as the summed probability distributions, a proxy for the radial density distribution of the low-inclination source KBOs and the newly released dust particles, are shown in Figure \ref{fig:Kepler}.

 For each KBO we have randomly generated a mean anomaly, $M_i$,  and calculated a resulting eccentric anomaly, $E_i$,  by solving Kepler's equation, $M_i = E_i - e_i \sin(E_i)$ \citep{danby:92a} to pick a position $r_i$ where a dust particle is assumed to be released.  The new radiation-pressure-induced orbital elements, $a_i^*, e_i^*$ can be calculated using equations \ref{eq:a} and \ref{eq:e}.  Figure \ref{fig:Kepler} shows the distributions of these new orbital elements, as well as the initial radial density distribution of the newly released dust particles for $\beta = 0.3$ and 0.4. Larger $\beta$ values result in increasing  $a^*$ and $e^*$ values, hence a more extended initial dust cloud, and a decreasing number of particles that remain bound in the solar system.

 \begin{figure}
     \centering
     \includegraphics[width=\linewidth]{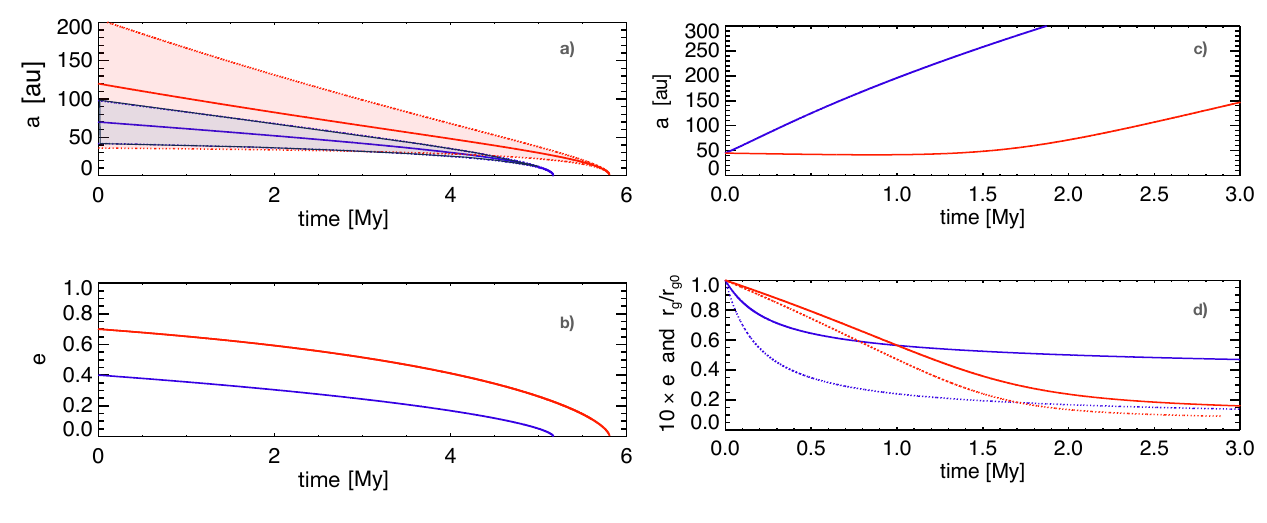}
     \caption{\textbf{Left:} Temporal variation due to PR drag of the semi-major axis \textbf{(a)} and eccentricity \textbf{(b)}  of olivine type particles with $\beta=0.3$ starting with initial $a=70$ au and $e = 0.4$ (blue lines) and $\beta = 0.4$ with initial $a =120$ au and $e = 0.7$ (red). These initial conditions were motivated by the peak of the $a$ and $e$ distributions as illustrated in Figure \ref{fig:Kepler}. The shaded regions show the range of their radial motion $a(1-e) \le r \le a(1+e)$. 
     \textbf{Right:} Temporal variation over 2 Myrs of the semi-major axis \textbf{(c)},  eccentricity (\textbf{d)} dotted lines), and radius (\textbf{d)} bottom solid lines) of 1 (blue line) and 5 $\mu$m (red) photo-sputtered ice particles, launched with an initial semi-major axis $a = 45$ au and eccentricity $e =0.1$. Small ice grains continuously gain energy and drift outwards out of the solar system, while bigger particles initially drift inward until their mass loss results in a large enough $\beta$ to reverse the direction of their drift, and eventually also leave the solar system.}
     \label{fig:combined}
 \end{figure}

 The subsequent orbital evolution of the released dust particles is driven by PR drag \citep{wyatt:50a, burns:79a}

\begin{eqnarray}
\frac{da}{dt} & = & -c\  \frac{\beta}{a}\ \frac{(2+3 e^2)}{(1-e^2)^{3/2}} 
\label{eq:ae1}\\
\frac{de}{dt} & = & -c\  \frac{\beta}{a^2}\  \frac{5e}{2(1-e^2)^{1/2}},
\label{eq:ae2}
\end{eqnarray}
where $c = 4.44\times 10^{15}$ cm$^2$/s is a numerical constant. Figure \ref{fig:combined} shows example histories of the orbital evolution of olivine type dust particles, indicating that the typical lifetime of small $\mu$m-sized particles in the KB is on the order of a few $\times 10^6$ years. 

Ferromagnesian olivines, pyroxenes, metal sulfides, and amorphous carbon are thought to dominate the mineral assemblage of IDPs in the solar system \citep{lisse:07a, zolensky:08a}. However, KBOs also contain a large amount of stable ices \citep{lisse:21a} that are likely expressed in dust grains emitted from them. Ice grains have higher $\beta$ values due to their low absorptivity and density but high reflectivity, and are susceptible to relatively rapid destruction due to photo-sputtering \citep{grigorieva:07a}. The photo-sputtering rate of icy particles is estimated to be 
\begin{equation}
\frac{d r_g}{dt}  \simeq 0.4/d^2 \ \ \ {\rm cm/My}, 
\label{eq:erode}
\end{equation}
where $r_g$ is the radius of the dust grain, and $d$ is the distance from the Sun in units of au \citep{harrison:67a, carlson:80a}. For grains with radii $r_g \ge 0.5 \  \mu$m, $\beta \simeq 0.5 \ r_{g0} /r_g$, where $r_{g0}$ is the dust grain radius where $\beta = 0.5$. For ice particles, $\beta = 0.5$ at $r_{g0} \simeq 5\ \mu$m  \citep{grigorieva:07a}. The expected photo-sputtering and, for comparison,  the Poynting-Robertson lifetimes of ice grains, setting $e \ll 1$, is \citep{burns:79a} 

\begin{eqnarray}
\tau_{PS}\  [{\rm yr}]& = & \frac{r_g}{ {\dot{r_g}}} = 250\ [{\rm yr}]\ r_g\  [\mu{\rm m}] 
 \ ( d\ [{\rm au}])^2,\\ 
 \tau_{PR}\  [{\rm yr}] & = & 400\ [{\rm yr}] \  ( d\ [{\rm au}])^2 /\beta = 400\  [{\rm yr}] \  ( d\ [{\rm au}])^2 / (r_{g0}/2 r_{0}), 
\end{eqnarray}
indicating that $\tau_{SP} \ll \tau_{PR}$ for small particles ($\beta < 0.5$). For example, a 1 $\mu$m radius ice particle  released at 40 au on a circular Kepler orbit experiences $\tau_{PS} \simeq 4 \times 10^4$  years. However, as small particles drift outward, their lifetimes rapidly increases and photo-sputtering will not result in total destruction, but rather ejection from the solar system. 

To investigate the combined effects of erosion and drag, we combine numerical time-step integrations of equations \ref{eq:ae1} and \ref{eq:ae2}, followed by an update of $r_g$ using equation \ref{eq:erode}, update $a$ and $e$ using equations  \ref{eq:a} and \ref{eq:e} with $\beta \rightarrow \Delta\beta$, and replace $r$ with its temporal average $<r> = a(1+0.5 e^2)$. Similarly, for  estimating an orbit-averaged erosion rate, we set $<r^{-2}> = \left(a^2(1+1.5e^2)\right)^{-1}$. Figure \ref{fig:combined} shows example histories of the orbital elements of eroding ice particles.
The resulting increase in $\beta$ of a quickly eroding grain sets up a competition between inward drift due to PR drag, outward drift due to increasing $\beta$, and, ultimately, ejection from the solar system. Hence, it might be the case that New Horizons is now encountering a new population of icy particles that current models, based on previous SDC measurements
\citep{poppe:16c}, could not have predicted. 

%Finally, optical surveys of distant KBOs are starting to discover KBOs beyond 55 au \citep{fraser:23a}. We cannot currently make any confident statements on the distribution of potentially dust producing bodies at large heliocentric distances, however, as these surveys progress, new populations may be revealed.

Finally,  our current understanding of dust-producing parent bodies in the KB is likely to be incomplete. Current optical surveys indicate that the KBOs' spatial distribution could extend further,  far beyond our current estimates \citep{fraser:23a}.

\FloatBarrier
%%%%%%%%%%%%%%%%%%%%
\section{Conclusion} \label{sec:conclusions}

New data (Figure \ref{flux}) at larger heliocentric distances show that the SDC reported dust fluxes, out to 55 au, remained higher than current models expected. Each of the three physical phenomena: optical properties and radiation pressure \citep{arnold:19a}; compositional variation between silicates and ice \citep{grigorieva:07a}; and a further reaching than currently identified distribution of dust-producing bodies in the solar system \citep{fraser:23a}, may contribute to the data/model deviation. The continued measurements identifying either increasing, constant,  or declining dust fluxes, combined with more detailed numerical investigations will help constrain the relative contributions from each of these mechanisms for dust production and transport in the outer KB.

New Horizons is expected to operate through the 2040s, exploring heliocentric distances beyond 100 au. The continued operation of SDC provides an opportunity to explore the outer edges of our solar system, and possibly record the transition into a new region in space where interstellar particles dominate the dust environment. Complementary to optical observations of the KB, SDC measurements provide a unique opportunity to learn more about our Kuiper Belt's extent, dust sources and populations beyond it, interstellar dust, and dust disks around other stars.

\begin{acknowledgments}
The authors acknowledge support from NASA's New Horizons mission, and useful discussions with Drs P. Pokorn\'y, and J.R. Szalay.
\end{acknowledgments}

\bibliography{refs_2023}{}
\bibliographystyle{aasjournal}

\end{document}